\documentclass[prd,preprint,superscriptaddress,showpacs,byrevtex]{revtex4} 
\usepackage{epsfig} 

\newcommand{\be}{\begin{equation}} 
\newcommand{\ee}{\end{equation}} 
\newcommand{\bea}{\begin{eqnarray}} 
\newcommand{\eea}{\end{eqnarray}} 
\def\simgt{\rlap{\lower 3.5 pt \hbox{$\mathchar \sim$}} \raise 1pt 
Ê \hbox {$>$}}

\begin{document} 

\begin{flushright}
YITP-SB-06-02
\end{flushright}

\title{Higgs Boson Production from Black Holes at the LHC} 

\author{Gouranga C. Nayak} \email{nayak@insti.physics.sunysb.edu} 
\affiliation{ C. N. Yang Institute for Theoretical Physics, 
Stony Brook University, SUNY, Stony Brook, NY 11794-3840, USA } 

\author{J. Smith} \email{smith@insti.physics.sunysb.edu} 
\affiliation{ C. N. Yang Institute for Theoretical Physics, 
Stony Brook University, SUNY, Stony Brook, NY 11794-3840, USA } 

\date{\today} 

\begin{abstract} 
If the fundamental Planck scale is near a TeV, then TeV scale black 
holes should be produced in proton-proton collisions at the LHC
where $\sqrt{s}$ = 14 TeV.  As the temperature of the black holes can 
be $\sim$ 1 TeV we also expect production of Higgs bosons from them 
via Hawking radiation.  
This is a different production mode for the Higgs boson,
which would normally be produced via direct pQCD parton fusion processes.
In this paper we compare total cross sections and 
transverse momentum distributions 
$d\sigma/dp_T$ for Higgs production from black holes at the LHC 
with those from direct parton fusion processes at next-to-next-to-leading
order and next-to-leading order respectively. We find that the 
Higgs production from black holes can be larger or smaller than the direct 
pQCD production depending upon the Planck mass and black hole mass.
We also find that $d\sigma/dp_T$ of Higgs production from black holes
increases as a function of $p_T$ which is in sharp contrast with the 
pQCD predictions where $d\sigma/dp_T$ decreases 
so we suggest that the measurement of an increase in $d\sigma/dp_T$ 
as $p_T$ increases for Higgs (or any other heavy particle) production can 
be a useful signature for black holes at the LHC.

\end{abstract} 
\maketitle 

\newpage 

\section{Introduction} 

It is now generally accepted that the scale of quantum gravity {\it could be} 
as low as one TeV \cite{folks}
and hence there can be graviton, radion and black hole production at LHC
\cite{ppbf,pp,pp1,pp2,pp3,ag,ppch,ppk,ppu,park,hof,more,gram,cham,pp4,pp5,pp6}. 
If such processes occur then LHC collider experiments \cite{gravr,gravr1} 
can probe TeV scale quantum gravity.  
One of the most exciting aspects of this 
will be the production of black holes in particle accelerators. These 
`brane-world' black holes will be our first window into the extra 
dimensions of space predicted by string theory, and required by the 
several brane-world scenarios that provide for a low energy 
Planck scale \cite{large}. As the black hole
masses at the LHC are relatively small (3-7 TeV) and the temperatures of the 
black holes are very high ($\sim$ 1 TeV) the black holes can be a source for 
Higgs boson production via Hawking radiation. 
In fact there can be an enormous amount of heavy (SUSY) particle 
production from black holes \cite{susy}, much more than expected from
normal pQCD processes \cite{been}. 
This comes about from two competing effects as the Planck scale 
increases: 1) Higgs production from black holes increases 
because the temperature of the black holes increases as the 
Planck scale increases for fixed black hole masses 
(see below) and 2) the cross section for black hole 
production decreases \cite{gray4,bv,grayd,cooper,plasma}.
Recently phenomenological analyses have been made to connect
the theoretical models with future data at the LHC,
\cite{YR,ravindran,allanach}.
Programs have been written to interface the theoretical
predictions to Monte Carlo generators for specific detectors 
such as D0 and CDF at Fermilab and ATLAS
at the LHC \cite{landsberg,harris}.     
Reviews of this exciting field are given in \cite{reviews}.  

In this paper we compare Higgs production cross sections
from TeV scale black hole production at the LHC via Hawking radiation 
with the direct pQCD parton fusion processes at 
next-to-next-to-leading order (NNLO) \cite{rsvn,others}.
After all if the Planck and black hole masses are larger than the
tentative estimates in the literature then extra dimensional models
may not yield any signals whatsoever at the LHC. Or the masses may
be so large that the signals from them are very small.
Therefore it is necessary to compare the standard pQCD
results for Higgs production with the corresponding black hole results. 
We find that the Higgs production cross sections from 
black holes at the LHC can be larger or smaller than those
from pQCD processes depending on the value of the TeV scale 
Planck mass and the black hole masses. We find that as long as the 
temperature of the black holes is of the order of one TeV, 
the Higgs production cross section from the 
black holes does not depend very much on the Higgs mass ($M_H$). 
On the other hand the direct pQCD production cross section 
is sensitive to $M_H$. 
This provides us with an important conclusion:  if TeV scale black holes 
are indeed formed at the LHC, then one signature of this 
will be an unusually copious production of massive (Higgs and
SUSY) particles, which 
is not possible via pQCD processes. Hence if we observe 
very high rates for massive particle production at the LHC, this might provide 
indirect evidence that TeV scale black holes are being produced. 

We also study the $p_T$ differential cross sections for Higgs production 
from black holes and from the pQCD parton fusion processes,
here in next-to-leading order (NLO) \cite{smith,diff1}.
One of the interesting results we find is that as long as the
temperature of the black holes is very high ($\sim$ 1 TeV) then 
$d\sigma/dp_T$ increases as $p_t$ increases 
(up to about $p_T$ equal to 1 TeV then it decreases). 
This is in sharp contrast to pQCD predictions
where $d\sigma/dp_T$ decreases as $p_T$ increases for fixed Higgs boson
masses. This is not only true for Higgs particles but also for any 
heavy (SUSY) particles \cite{susy} emitted from a black hole via 
Hawking radiation 
as long as the temperature of the black hole is high ($\sim$ 1 TeV). 
Therefore if one experimentally observes that
the $d\sigma/dp_T$ for heavy final state particles
increases as $p_T$ increases (up to about 1 TeV or higher)
then it might provide a good evidence for black hole production at the LHC. 

The paper is organized as follows.  In Sec. II we present the  
computation for the rate of Higgs production 
and its transverse momentum distribution 
from black holes via Hawking radiation at the LHC.
In Sec. III we sketch the calculation of the pQCD 
total and $p_T$ differential cross sections for Higgs production at the LHC.
In Sec. IV we present and discuss our results.

\section{Higgs boson production from black holes at the LHC } 

If black holes are formed at the LHC then they will quickly evaporate 
by emitting thermal Hawking radiation. The emission rate per unit time 
for a Higgs particle with momentum $p =|\vec p|$ and 
energy $Q= \sqrt{p^2+M_H^2}$ can be written \cite{gram} as 
\be 
\frac{dN}{dt }= 
\frac{c_s \sigma_s}{8\pi^2}\frac{dp\, p^2}{(e^{Q/T_{BH}} - 1)}\,, 
\label{thermal}
\ee 
where $\sigma_s$ is the grey body factor and $T_{BH}$ is the black hole 
temperature, which depends on the number of extra dimensions and on 
the TeV scale Planck mass. 
$c_s$ is the multiplicity factor.
The temperature of the black hole is given in \cite{pp}, namely 
\be 
T_{BH}=\frac{d+1}{4\pi R_{S}} ~=~
\frac{d+1}{4 \sqrt{\pi}}~
M_P [\frac{M_P}{M_{BH}}\frac{d+2}{8\Gamma(\frac{d+3}{2})}]^{\frac{1}{1+d}}\,, 
\ee 
where $R_{S}$ is the Schwarzschild radius of the black hole, 
$M_P$ is the TeV scale Planck mass, $M_{BH}$ is the mass of the black hole
and $d$ is the number of extra dimensions. 
The grey body factor in the geometrical approximation is 
given by \cite{gray4,bv,grayd}
\be 
\sigma_s = \Gamma_s 4\pi (\frac{d+3}{2})^{2/(d+1)} \frac{d+3}{d+1} R^2_{S}\,, 
\label{sigm}
\ee 
where we take $\Gamma_s$ = 1 for scalars. Recent work \cite{harris}
shows that grey body factors for scalar emission do not vary much with $d$
in contrast to fermion and gauge boson emission
so Eq. ({\ref {sigm}}) is a reasonable ansatz. 
From Eq. ({\ref {thermal}}) we get: 
\be 
\frac{dN}{dt dp}= \frac{c_s \sigma_s}{8\pi^2}
\frac{p^2 }{(e^{\sqrt{p^2+M_H^2}/T_{BH}} - 1)}\,. 
\ee 
The total number of Higgs particles emitted from the black holes is thus given by: 
\be N_{Higgs}= \int_0^{t_f} dt \int_0^{M_{BH}} dp\, 
\frac{c_s \sigma_s}{8\pi^2}\frac{p^2 }{(e^{\sqrt{p^2+M_H^2}/T_{BH}} - 1)}\,, 
\label{NN}
\ee 
where $t_f$ is the total time taken by the black hole to completely evaporate, 
which takes the form \cite{pp1}:
\be
t_f~=~\frac{C}{M_P}(\frac{M_{BH}}{M_P})^{\frac{d+3}{d+1}}\,.
\ee
$C$ depends on the extra dimensions and on the polarization degrees 
of freedom, etc. 
However, the complete determination of $t_f$
depends on the energy density present outside the black hole
which is computed in \cite{cooper} where 
the absorption of the quark-gluon plasma \cite{plasma} 
by a TeV scale black hole at the LHC is considered
(this time is typically about $10^{-27}$ sec). 
The value we use throughout this paper is $t_f$= $10^{-3}$ fm
which is the inverse of the TeV scale energy. 

This result in Eq. ({\ref{NN}}) is for Higgs particle emission from 
black holes of temperature $T_{BH}$. 
To obtain the Higgs production cross section from all 
black holes produced in proton-proton collisions at the LHC 
we need to multiply the black hole 
production cross section with the number of Higgs bosons produced from a 
single black hole. 
The black hole production cross section 
$\sigma_{BH} $ in high energy hadronic collisions at zero 
impact parameter is given in \cite{pp,cham}, namely 
\bea 
\sigma_{BH}^{AB \rightarrow BH +X}(M_{BH}) 
= {\sum}_{ab}~ 
\int_{\tau}^1 dx_a \int_{\tau/x_a}^1 dx_b f_{a/A}(x_a, \mu^2) 
\nonumber \\ 
\times f_{b/B}(x_b, \mu^2) 
\hat{\sigma}^{ab \rightarrow BH }(\hat s) ~\delta(x_a x_b -M_{BH}^2/s). 
\label{bkt} 
\eea 
In this expression $x_a (x_b)$ is the longitudinal momentum 
fraction of the parton inside 
the hadron A(B) and $\tau=M_{BH}^2/s$, where $\sqrt s$ is the 
hadronic center-of-mass energy. 
Energy-momentum conservation implies $\hat s =x_ax_b s=M_{BH}^2$. 
We use $\mu = M_{BH}$ as the scale at which the 
parton distribution functions 
$f_{a/A}, f_{b/B}$ are measured. ${\sum}_{ab}$ represents 
the sum over all partonic contributions.
The black hole production cross section in a binary partonic 
collision is given by \cite{pp} 
\bea 
\hat{\sigma}^{ab \rightarrow BH }(\hat s) = 
\frac{1}{M^2_P} [\frac{M_{BH}}{M_P}
(\frac{8\Gamma(\frac{d+3}{2})}{d+2})]^{2/(d+1)}\,, 
\label{bk3} 
\eea 
where $d$ denotes the number of extra spatial dimensions. Note that $M_{BH}$
should be approximately five times $M_P$ for the classical limit to
apply and Hawking evaporation to occur. We choose $M_{BH} = 3 M_P$
and $M_{BH} = 5 M_P$ for our plots. 
The total cross section for Higgs production at LHC is then given by 
\be 
\sigma_{Higgs} = N_{Higgs} \sigma_{BH}. 
\label{susybk}
\ee 
We will compare this cross section for Higgs boson
production via black hole resonances with the Higgs cross section 
produced via pQCD processes, as will be explained in the next 
section. To compare
the differential cross sections we decompose the phase-space integration
in Eq. (\ref{thermal}) 
as $d^3\vec p ~=~d^2p_t ~dp_z~=~d^2p_t~m_t~\cosh y~dy$
where 
$p^{\mu}~=~( \sqrt{p_t^2+M_H^2} \cosh y, p_x,p_y, \sqrt{p_t^2+M_H^2} 
\sinh y)$ and integrate over the rapidity $y$.

\section{Direct Higgs Production in pp Collisions at the LHC} 

The LEP experiments \cite{lep} give a lower mass limit on the mass of
the Higgs $M_H\sim 114~{\rm GeV/c^2}$ and fits to the data using 
precision calculations
in the electro-weak sector of the standard model indicate an upper limit 
$m_H<200~{\rm GeV/c^2}$ with $95~\%$ confidence level. Therefore we will
concentrate on the mass interval 
$100~{\rm GeV/c^2}  \le  M_H \le  300~{\rm GeV/c^2}$.  

At the LHC proton-proton collider the
dominant QCD production process involves the gluon-gluon fusion
mechanism. (We comment on the weak boson fusion reaction at the
end of the paper).
In the standard model the Higgs boson couples to the gluons
via heavy quark loops. 
Since the coupling of the scalar Higgs boson ${\rm H}$ to a fermion loop 
is proportional to the mass of the fermion (for a review see \cite{ghkd}), 
the top-quark loop is the most important.  
In lowest order (LO) the gluon-gluon fusion process 
$g + g \rightarrow {\rm H}$, represented by the top-quark triangle graph, 
was computed in \cite{wil}. 
The two-to-two body tree graphs, given by
gluon bremsstrahlung $g + g \rightarrow g + {\rm H}$, 
$g + q(\bar q) \rightarrow q(\bar q) + {\rm H}$ and
$q+\bar q \rightarrow g + {\rm H}$ were computed  
in \cite{ehsb}. From these reactions one can derive 
the transverse momentum ($p_T$) and rapidity ($y$) distributions of the
scalar Higgs boson. The total integrated
cross section, which also involves the computation of the QCD corrections 
to the top-quark loop, has been calculated in \cite{gsz}.
This calculation was rather cumbersome since it involved the computation
of two-loop triangular graphs with massive quarks. Furthermore also the 
two-to-three parton reactions have been computed in \cite{dkosz} using 
helicity methods. From the experience gained from the next-to-leading (NLO) 
corrections presented in \cite{gsz} it 
is clear that it will be very difficult to obtain the exact NLO corrections
to one-particle inclusive distributions as well as the NNLO corrections to
the total cross section.

Fortunately one can simplify the calculations if one takes the
large top-quark mass limit $m_t \rightarrow \infty$. 
In this case the Feynman graphs are obtained from an effective Lagrangian 
describing the direct coupling of the scalar Higgs boson
to the gluons. The LO and NLO contributions to the total cross section
in this approximation were computed in \cite{dawson}. A thorough analysis
\cite{gsz,kls} reveals that the error 
introduced by taking the 
$m_t\rightarrow \infty$ limit is less than about $5\%$ provided $m_H\le 2~m_t$. 
The two-to-three body processes were computed with the effective 
Lagrangian approach for the scalar Higgs bosons 
in \cite{kdr} using helicity methods. 
The one-loop corrections
to the two-to-two body reactions above were computed for the scalar
Higgs boson in \cite{schmidt}.
These matrix elements were used to compute the transverse 
momentum and rapidity distributions of the scalar Higgs boson up to NLO in
\cite{smith,fgk}.
The effective Lagrangian method was also applied to obtain the NNLO total 
cross section for scalar Higgs production by the calculation of the two-loop 
corrections to the 
Higgs-gluon-gluon vertex in \cite{harland,rsvn2}, 
the soft-plus-virtual gluon 
corrections in \cite{haki1} and the computation of the two-to-three 
body processes in \cite{others,smith}. 

In the large top-quark mass limit, which we use from now on,
the Feynman rules 
for scalar Higgs  production can be derived from the following 
effective Lagrangian density
\begin{eqnarray}
\label{eqn2.1}
{\cal L}^{\rm H}_{eff}=G_{\rm H}\,\Phi^{\rm H}(x)\,O(x) \quad 
\mbox{with} \quad O(x)=-\frac{1}{4}\,G_{\mu\nu}^a(x)\,G^{a,\mu\nu}(x)\,,
\end{eqnarray}
where $\Phi^{\rm H}(x)$ represents the scalar field. 
Furthermore the gluon field strength is given by $G_a^{\mu\nu}$.
The factor multiplying the operator is chosen in such a way that the 
vertices are normalised to the effective coupling constant $G_{\rm H}$. 
The latter is determined by the top-quark triangular loop graph, 
which describes the decay process ${\rm H} \rightarrow g + g$  
including all QCD corrections, taken in the
limit that the top-quark mass $m_t\rightarrow \infty$. This yields
\begin{eqnarray}
\label{eqn2.3}
G_{\rm H}&=&-2^{5/4}\,a_s(\mu_r^2)\,G_F^{1/2}\,
\tau_{\rm H}\,F_{\rm H}(\tau_{\rm H})\,{\cal C}_{\rm H}
\left (a_s(\mu_r^2),\frac{\mu_r^2}{m_t^2}\right )\,,
\end{eqnarray}
where $a_s(\mu_r^2)$ is defined by
\begin{eqnarray}
\label{eqn2.4}
a_s(\mu_r^2)=\frac{\alpha_s(\mu_r^2)}{4\pi}\,,
\end{eqnarray}
with $\alpha_s(\mu_r^2)$ the running coupling constant and $\mu_r$ 
the renormalization scale. Further $G_F$ represents the Fermi constant and the
function $F_{\rm H}$ is given by
\begin{eqnarray}
\label{eqn2.5}
&& F_{\rm H}(\tau)=1+(1-\tau)\,f(\tau)\,, \qquad  
\tau=\frac{4\,m_t^2}{M_H^2} \,, 
\nonumber\\[2ex]
&&f(\tau)=\arcsin^2 \frac{1}{\sqrt\tau}\,, \quad \mbox{for} \quad \tau \ge 1\,,
\nonumber\\[2ex]
&& f(\tau)=-\frac{1}{4}\left ( \ln \frac{1-\sqrt{1-\tau}}{1+\sqrt{1-\tau}}
+\pi\,i\right )^2 \quad \mbox{for} \quad \tau < 1\,,
\end{eqnarray}
In the large $m_t$-limit we have
\begin{eqnarray}
\label{eqn2.6}
 \mathop{\mbox{lim}}\limits_{\vphantom{\frac{A}{A}} \tau \rightarrow \infty}
F_{\rm H}(\tau)=\frac{2}{3\,\tau}\,. 
\end{eqnarray}
The coefficient function ${\cal C}_{\rm H}$ originates from the corrections 
to the top-quark triangular graph provided one takes the 
limit $m_t\rightarrow \infty$.  
The coefficent function has been computed 
up order $\alpha_s^2$ in \cite{kls,cks} for the Higgs. 

Using the effective Lagrangian approach the total cross section of the reaction
\begin{eqnarray}
\label{eqn2.10}
H_1(P_1)+H_2(P_2)\rightarrow {\rm H}(-p_5)+'X'\,,
\end{eqnarray}
where $H_1$ and $H_2$ denote the incoming hadrons and $X$ represents an 
inclusive hadronic state has been calculated to NNLO.
This total cross section is given by
\begin{eqnarray}
\label{eqn2.11}
&&\sigma_{\rm tot}=\frac{\pi\,G_{\rm H}^2}{8\,(N^2-1)}\,\sum_{a,b=q,\bar q,g}\,
\int_x^1 dx_1\, \int_{x/x_1}^1dx_2\,f_a(x_1,\mu^2)\,f_b(x_2,\mu^2)\,
\nonumber\\[2ex] && \qquad\qquad\times
\Delta_{ab,{\rm H}}\left ( \frac{x}{x_1\,x_2},\frac{M_H^2}{\mu^2} \right ) \,,
\nonumber\\[2ex]
&&\mbox{with}\quad x=\frac{M_H^2}{S} \quad\,,\quad S=(P_1+P_2)^2\quad
\,,\quad p_5^2=M_H^2\,,
\end{eqnarray}
where the factor $1/(N^2-1)$ originates from the colour average
in the case of the local gauge group SU(N). Further
we have assumed that the scalar Higgs boson is mainly
produced on-shell i.e. $p_5^2\sim M_H^2$. The parton densities denoted by
$f_a(y,\mu^2)$ ($a,b=q,\bar q,g$)
depend on the mass factorization/renormalization scale $\mu$. 
The same scales also enter the coefficient functions $\Delta_{ab,{\rm H}}$ 
which are derived from the partonic cross sections.

Up to NNLO we have to compute the following partonic subprocesses.
On the Born level we have the reaction
\begin{eqnarray}
\label{eqn3.2}
g+g \rightarrow {\rm H}\,.
\end{eqnarray}
In NLO we have in addition to the one-loop virtual corrections to the above
reaction the following two-to-two body processes
\begin{eqnarray}
\label{eqn3.3}
g+g \rightarrow {\rm H} + g \quad\,,\quad
g+q(\bar q) \rightarrow {\rm H} + q(\bar q)
\quad\,,\quad q+ \bar q \rightarrow {\rm H} + g\,.
\end{eqnarray}
In NNLO we receive contributions from the two-loop virtual corrections to
the Born process in Eq. (\ref{eqn3.2}) and the one-loop corrections to
the reactions in  Eq. (\ref{eqn3.3}). To these contribution one has to add
the results obtained from the following two-to-three body reactions
\begin{eqnarray}
\label{eqn3.4}
&&g+g \rightarrow {\rm H} + g + g \quad\,,\quad
 g+g \rightarrow {\rm H} + q_i + \bar q_i\,,
\\[2ex]
\label{eqn3.5}
&&g+q(\bar q) \rightarrow {\rm H} + q(\bar q) + g\,,
\\[2ex]
\label{eqn3.6}
&& q+ \bar q \rightarrow {\rm H} + g + g \quad\,,\quad
q+ \bar q \rightarrow {\rm H} + q_i + \bar q_i\,,
\\[2ex]
\label{eqn3.7}
&&q_1+ q_2 \rightarrow {\rm H} + q_1+ q_2 \quad\,,\quad 
q_1+ \bar q_2 \rightarrow {\rm H} + q_1+ \bar q_2\,, 
\\[2ex]
\label{eqn3.8}
&&q+ q \rightarrow {\rm H} + q+ q\,.
\end{eqnarray} 
The computation of the phase space integrals has been done in
\cite{rsvn}, to which we refer for further details (see also \cite{others}).
After they have been calculated the partonic cross section is
rendered finite by coupling constant renormalization, operator renormalization
(see \cite{klzu}) and the removal of collinear divergences.  
In the representation of the coefficient functions above we have
set the renormalization scale $\mu_r$ equal to the mass 
factorization scale $\mu$. 
The final total cross section 
for Higgs-boson production in proton-proton collisions at the LHC 
can be written as
\begin{eqnarray}
\label{eqn4.1}
\sigma_{\rm tot}=\frac{\pi\,G_{\rm H}^2}{8\,(N^2-1)}\,\sum_{a,b=q,\bar q,g}
\int_x^1\,dy\, \Phi_{ab}(y,\mu^2)\,
\Delta_{ab}\left (\frac{x}{y},\frac{M_H^2}{\mu^2}\right )\,,
\end{eqnarray}
where $x=M_H^2/S$ and $\Phi_{ab}$ is the parton-parton flux defined by
\begin{eqnarray}
\label{eqn4.2}
\Phi_{ab}(y,\mu^2)=\int_y^1\frac{du}{u}\,f_a(u,\mu^2)\,f_b\left (
\frac{y}{u},\mu^2\right )\,.
\end{eqnarray}
The coefficient functions $\Delta_{ab}$ in the effective 
Lagrangian approach were 
computed exactly in NNLO and the parton densities are also known to the
same order because the exact three-loop splitting
functions (anomalous dimensions) have now been calculated \cite{mvv}.
Hence we can check whether the approach of using only 
a finite number of moments (see \cite{larin2}) which was used in 
\cite{nevo} together with other constraints to approximate the splitting 
functions is accurate.
These approximations are very reliable as long as $y>10^{-4}$ in Eq. 
(\ref{eqn4.1}). The approximated splitting functions were used in 
\cite{mrst01} and \cite{mrst02} to obtain NNLO parton density sets. 
For the NLO and NNLO plots we employ the two-, and three-loop
asymptotic forms of the running coupling constant as given in Eq. (3) of
\cite{chkn}. For our plots we take $\mu=M_H$ and use the 
MRST set above for the NNLO computations. 
For the computation of the effective coupling constant $G_{\rm H}$ in Eq. 
(\ref{eqn2.3}) we choose the top quark mass $m_t=173.4~{\rm GeV/c^2}$ and the
Fermi constant $G_F=1.16639~{\rm GeV}^{-2}=4541.68~{\rm pb}$. 

We now discuss briefly the calculation of the $p_T$ differential cross 
section for Higgs boson production in proton-proton collisions at the LHC.
This was done by a different method since we only integrated analytically
over part of final phase space. 
We still need the 2 to 2 parton fusion processes 
$g + g \rightarrow g + H$,
$q + \bar q  \rightarrow g + H$,
and $q(\bar q) + g \rightarrow q(\bar q) + H$,
together with all the real (2 to 3) and virtual NLO corrections \cite{smith},
\cite{diff1}.
The $p_T$ differential cross section for Higgs production 
in $H_1(P_1) + H_2(P_2) \rightarrow H(-p_5) + 'X'$ at NLO is given by
\begin{eqnarray}
\label{eqn6.14}
\frac{d~\sigma}{d~p_T}= \int_{-y_{{\rm max}}}^{y_{{\rm max}}} dy
\,\frac{d^2~\sigma^{{\rm H_1H_2}}}{d~p_T~d~y}
(S,p_T^2,y,M_H^2)\,,
\end{eqnarray}
where 
\begin{eqnarray}
\label{eqn6.10}
S \frac{d^2~\sigma^{{\rm H_1H_2}}}{d~p_T^2~d~y}(S,p_T^2,y,M_H^2)=
S^2 \frac{d^2~\sigma^{{\rm H_1H_2}}}{d~T~d~U}(S,T,U,M_H^2)\,.
\end{eqnarray}
$M_H$ is the mass of the Higgs boson and $y$ is its rapidity.
The hadronic kinematical variables are defined by
\begin{eqnarray}
\label{eqn6.2}
S=(P_1+P_2)^2 \,, \qquad T=(P_1+p_5)^2\,, \qquad U=(P_2+p_5)^2 \,.
\end{eqnarray}
The 2-2 parton momenta satisfy $p_1 + p_2 + p_3 + p_5 =0$ in LO.
The invariants are given by
\begin{eqnarray}
\label{eqn6.9}
&&T=M_H^2-\sqrt S\,\sqrt{p_T^2+M_H^2}\,\cosh y+\sqrt S\,\sqrt{p_T^2+M_H^2}
\,\sinh y\,,
\nonumber\\[2ex]
&&U=M_H^2-\sqrt S\,\sqrt{p_T^2+M_H^2}\,\cosh y-\sqrt S\,\sqrt{p_T^2+M_H^2}
\,\sinh y\,.
\end{eqnarray}
The rapidity interval is given by
$ -y_{\rm max} \le y \le y_{\rm max} $ where
\begin{eqnarray}
\label{eqn6.13}
y_{{\rm max}}=
\frac{1}{2}\ln\frac{1+\sqrt{1-sq}}{1-\sqrt{1-sq}}\,,
\qquad sq=\frac{4~S~(p_T^2+M_H^2)}{(S+M_H^2)^2}\,.
\end{eqnarray}

The hadronic cross sections $d\sigma^{\rm H_1H_2}$ are related to the 
partonic level cross sections $d\sigma_{ab}$ as follows
\begin{eqnarray}
\label{eqn6.1}
S^2 \frac{d^2~\sigma^{{\rm H_1H_2}}}{d~T~d~U}(S,T,U,M_H^2)&=& \sum_{a,b=q,g}
\int_{x_{1,{\rm min}}}^1 \frac{dx_1}{x_1} \int_{x_{2,{\rm min}}}^1 
\frac{dx_2}{x_2}\,
f_a^{\rm H_1}(x_1,\mu^2)
\nonumber\\[2ex]
&&\times f_b^{\rm H_2}(x_2,\mu^2)\,s^2 
\frac{d^2~\sigma_{ab}}{d~t~d~u} (s,t,u,M_H^2,\mu^2)\,,
\end{eqnarray}
where $\mu$ is the factorization scale in the parton densities,
chosen to satisfy $\mu^2 = p_T^2 + M_H^2$.
In the case parton $p_1$ emerges from hadron $H_1(P_1)$ and parton
$p_2$ emerges from hadron $H_2(P_2)$ 
\begin{eqnarray}
\label{eqn6.3}
&& p_1=x_1\,P_1\,, \qquad p_2=x_2\,P_2 \,,
\nonumber\\[2ex]
&& s=x_1\,x_2\,S \,, \quad t=x_1(T-M_H^2)+M_H^2 \,, 
\quad u=x_2(U-M_H^2)+M_H^2\,,
\nonumber\\[2ex]
&& x_{1,{\rm min}}=\frac{-U}{S+T-M_H^2}\,, \qquad
x_{2,{\rm min}}=\frac{-x_1(T-M_H^2)-M_H^2}{x_1S+U-M_H^2}\,.
\end{eqnarray}
Further details on the calculation of the 2 to 2 and 2 to 3 partonic
cross sections are available in \cite{smith}, (see also \cite{diff1}).
We have computed the all the differential cross sections using the CTEQ6M
NLO parton density set in \cite{lai}.

\section{Results and Discussions} 

In this section we will compute Higgs production cross sections 
and $p_T$ distributions at $\sqrt{s}$ = 14 TeV in pp collisions. 
We begin with black hole production. 
We choose the factorization and normalization scales to be equal to
the mass of the black hole which is of the order of one TeV
and use CTEQ6M parton density distributions \cite{lai}. 
The computation of the black hole 
production cross sections follows from Eqs.(7) and (8) in Sec. II.

\begin{figure}[htb]
\vspace{2pt}
\centering{\rotatebox{270}{\epsfig{figure=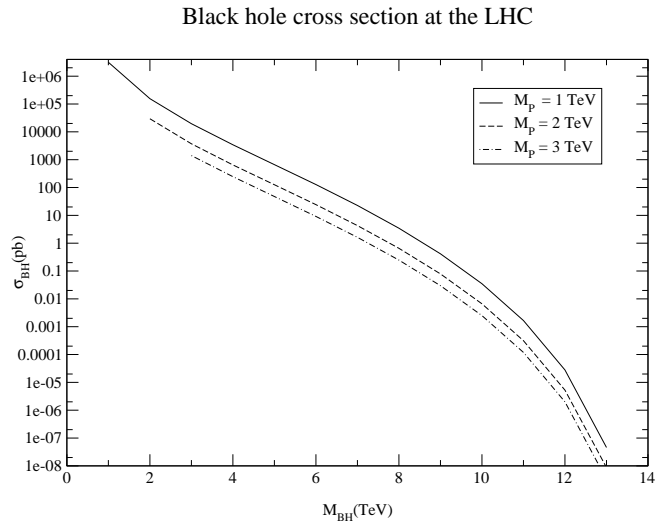,height=10cm}}}
\caption{ Total cross sections for black hole production at the LHC.}
\label{fig1}
\end{figure}

In Fig. 1 we plot the black hole production cross section
$\sigma_{BH}$ in pb at the LHC as a function of the 
black hole mass $M_{BH}$ in TeV.
The solid, dashed and dot-dashed curves are for 
Planck masses of 1, 2 and 3 TeV respectively. 
The number of extra dimensions $d=4$. As can be 
seen from the figure the cross sections decrease rapidly when
both the Planck and black hole masses increase. These
black hole production cross sections will be 
multiplied with the number of Higgs bosons produced from a 
single black hole to obtain the Higgs production cross section
from a black hole at the LHC.

As the temperature of the black hole at the LHC
is $\sim$1 TeV there is not much difference in the Higgs
production cross section from black holes if $M_H$ 
is increased from 120 to 200 GeV. Hence we will use
$M_H= $ 200 GeV when we compute the Higgs production
differential and total cross sections from the black holes.
In Fig.2 we present the total Higgs boson production cross section
from black hole production (from Eq.(9) in Sec.II) and from pQCD in NNLO. 
The former is given for two different choices of the Planck mass, each with
two choices of the black hole mass, namely
$M_P= 1$ TeV with $M_{BH}= 3,5 $ TeV and 
$M_P= 2$ TeV with $M_{BH}= 6,10 $ TeV. 
We plot for comparison the NNLO Higgs boson cross section from \cite{rsvn} 
as a function of the mass of the Higgs boson in GeV. 
For the latter curve we use the NNLO MRST parton 
density set in \cite{mrst01} because there is no NNLO CTEQ set.
We concentrate on the mass range 
100 GeV $\le M_H \le$ 300 GeV, which is where the bounds from the
LEP data indicate that it should be. We see that the Higgs production rate
from black holes can be larger or smaller than the direct pQCD
production depending upon the values of the Planck mass and the 
black hole mass. 

\begin{figure}[htb]
\vspace{2pt}
\centering{\rotatebox{270}{\epsfig{figure=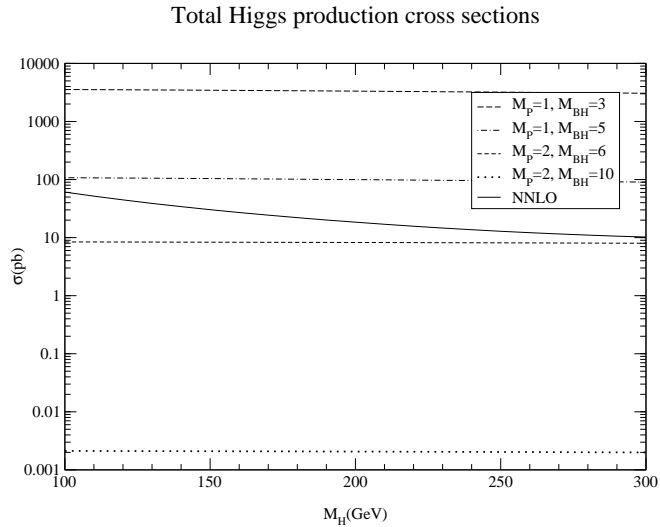,height=10cm}}}
\caption{ Total cross sections for Higgs production
from black holes and from direct pQCD processes at NNLO.}
\label{fig2}
\end{figure}

In Fig. 3 we present results for the $p_T$ differential cross 
sections in pb/GeV for Higgs production both via direct parton
fusion processes at NLO and indirect Higgs emission through black hole 
production (we integrate over the rapidity using the
transformation described after Eq.(9)).
In this figure we use CTEQM6 parton densities in \cite{lai}.
The three decreasing lines (as $p_T$ increases) are for
NLO parton fusion processes, with the solid, dashed and dot-dashed
lines for Higgs masses equal to 120, 160 and 200 GeV respectively.  
The three increasing lines (as $p_T$ is increased) are from
emissions from black hole production, with the solid, dashed and
dot-dashed lines for black hole masses equal to 3, 4 and 
5 TeV respectively with the Planck mass equal to 1 TeV in 
each case. 

\begin{figure}[htb]
\vspace{2pt}
\centering{\rotatebox{270}{\epsfig{figure=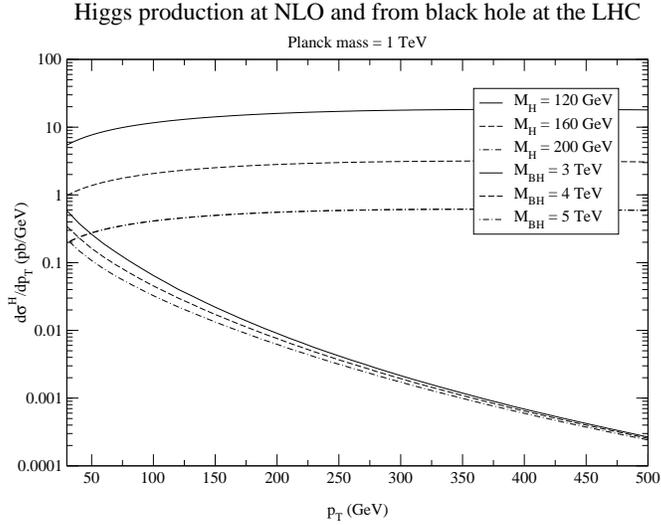,height=10cm}}}
\caption{ $p_T$ differential cross sections for Higgs production
from black holes and from direct pQCD processes at NLO.
The Planck mass is 1 TeV.}
\label{fig3}
\end{figure}

It can be seen from Fig. 3 that, with this choice of masses,
the $p_T$ differential
cross section for Higgs production from black hole emission
is larger than that from the direct NLO parton fusion processes
when $p_T \ge 50 $ GeV.
This is interesting because if a black hole
is indeed formed and if the Planck scale is around a TeV
then we may get more Higgs production from the black hole
than that we would have obtained from direct parton fusion processes.
This will enhance the chance of detecting Higgs bosons at the LHC.

It can also be seen that $d\sigma/dp_T$ for Higgs production from black holes
increases as $p_T$ increases whereas it decreases in the case of the
pQCD NLO calculation. This is a unique feature of 
black hole production at the LHC. The reason for this is that the
mass of the black hole formed is quite small ($M_{BH} \sim$ 5 TeV)
and hence the temperature of the black hole is very large
$\sim$ 1 TeV. For Higgs production with such a high temperature
the Bose-Einstein distribution function $e^{\sqrt{p^2+M_H^2}/T_{BH}} - 1$ 
in Eq. (\ref{NN}) remains almost flat with
respect to $p_T$ as long as $p_T$ is not much larger than
$T_{BH}$. Hence the increase 
of $d\sigma/dp_T$ as $p_T$ inceases comes from the 
increase in the transverse momentum phase space 
as can be seen from Eq. (\ref{NN}).

\begin{figure}[htb]
\vspace{2pt}
\centering{\rotatebox{270}{\epsfig{figure=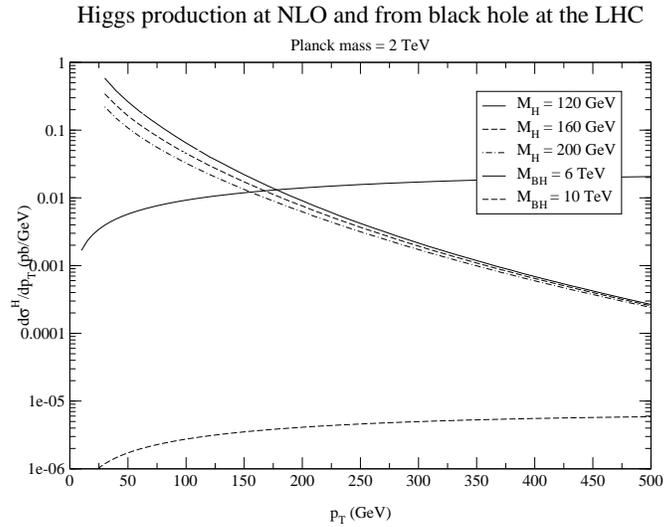,height=10cm}}}
\caption{ $p_T$ differential cross sections for Higgs production
from black holes and from direct pQCD processes at NLO.
The Planck mass is 2 TeV.}
\label{fig4}
\end{figure}

In Fig. 4 we present similar results as in the case
of Fig. 3 but now for a Planck mass of 2 TeV. 
The three decreasing lines (as $p_T$ increases) are from the 
NLO parton fusion processes, with the solid, dashed and dot-dashed 
lines for Higgs masses of 120, 160 and 200 GeV respectively.  
The two increasing lines (as $p_T$ increases) are from
black hole emissions, with the solid and dashed lines
for black hole masses equal to 6 and 10 TeV respectively. 
Clearly the differential cross section is quite small when $M_{BH}=10$ TeV
because the total cm energy is only 14 TeV.

The results in Figs. 3 and 4 show that as the Planck mass increases the 
Higgs production rate from black hole emission becomes smaller than that
from the NLO pQCD Higgs production. However as the $p_T$
is increased then at some value the Higgs production from black hole emission
will be larger than that of the direct pQCD Higgs production. Therefore 
the cross section from the emision of the black hole via
Hawking radiation will dominate over any other standard
model processes for large masses and large enough $p_T$. 
Hence a large rate of particle production at the LHC 
at high $p_T$ and high mass can be a 
possible signature of black hole production.

We have concentrated on Higgs production via QCD reactions and ignored
possible contributions from the so-called weak boson fusion reactions
involving the couplings of W and Z bosons to the Higgs.
In the latter case a typical partonic reaction is 
\begin{eqnarray}
q + q \rightarrow  {\rm H} + q + q\, 
\end{eqnarray}
where two virtual Z-bosons are exchanged between the quarks and the Higgs.
This is a typical t-channel process.  Virtual $W^+$ and $W^-$ bosons can 
also be exchanged.  The reason we have neglected this so-called WW/ZZ 
reaction is that it is not the dominant
contribution to {\it inclusive} Higgs production at large $p_T$. We show this
in Fig.5 where the contribution from the weak boson fusion reaction
was provided by J. Campbell from his NLO calculation in \cite{bc}.
Here $M_H = 120$ GeV and CTEQ6 parton densities have been used. Note that 
the scale is now in fb/GeV.
If one identifies the two final state jets and applies $p_T$ cuts on them 
then it is possible to enhance the weak boson signal and decrease the
QCD signal. However this now involves the study of exclusive
processes which is not the subject of this paper.

\begin{figure}[htb]
\vspace{2pt}
\centering{\rotatebox{270}{\epsfig{figure=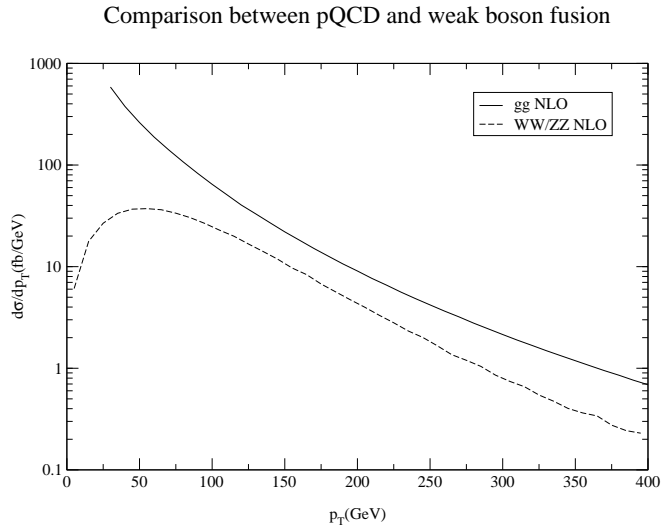,height=10cm}}}
\caption{ $p_T$ differential cross sections for Higgs production at the 
LHC from direct pQCD processes at NLO and weak boson fusion processes at NLO.}
\label{fig5}
\end{figure}

In summary, we have compared Higgs production total and
differential cross sections in proton-proton collisions
at the LHC ($\sqrt{s}$ = 14 TeV)  via pQCD processes 
and via Hawking radiation from black holes within the model
of TeV scale gravity. As the temperature 
of the black hole is $\sim$ 1 TeV there is a huge amount of Higgs 
production from black holes at the LHC if the Planck mass is $\sim$ 1 TeV.
We also find that $d\sigma/dp_T$ for 
Higgs production increases as a function of $p_T$ 
in sharp contrast with the pQCD predictions where it decreases. 
Hence we suggest that the measurement of an increase in $d\sigma/dp_T$ for
any heavy (Higgs or SUSY) particle production at the LHC as $p_T$ increases 
can be a useful signature for black hole production.

\acknowledgments

This work was supported 
in part by the National Science Foundation, grants PHY-0071027, 
PHY-0098527, PHY-0354776 and PHY-0345822. We thank J. Campbell for
providing the WW/ZZ NLO result in Fig. 5.


\begin{thebibliography}{aaaaaaa}

\bibitem{folks}
N. Arkani-Hamed, S. Dimopoulos and G.R. Dvali,
Phys. Lett. {\bf B429}, 263 (1998); Phys. Rev. D{\bf 59},086004 (1999);
I. Antoniadis, N. Arkani-Hamed, S. Dimopoulos and G.R. Dvali,
Phys. Lett. {\bf B436}, 257 (1998);
L.~Randall and R.~Sundrum, Phys. Rev. Lett. {\bf 83}, 3370 (1999);
{\bf 83}, 4690 (1999).

\bibitem{ppbf} T. Banks and W. Fischler, hep-th/9906038.

\bibitem{pp} S.~Dimopoulos and G.~Landsberg,
Phys.\ Rev.\ Lett.\  {\bf 87}, 161602 (2001); P.C. Argyres, S. Dimopoulos
and J. March-Russell, Phys. Lett. {\bf B441}, 96 (1998).

\bibitem{pp1} S.~B.~Giddings and S.~Thomas,
Phys.\ Rev.\ D {\bf 65}, 056010 (2002).

\bibitem{pp2} S.~B.~Giddings,
in {\it Proc. of the APS/DPF/DPB Summer Study on the Future of Particle Physics
(Snowmass 2001) } ed. R.~Davidson and C.~Quigg, hep-ph/0110127;
M.B. Voloshin, Phys. Lett.  {\bf B518}, 137 (2001); Phys. Lett.
{\bf B524}, 376 (2002).

\bibitem{pp3} 
D.~M.~Eardley and S.~B.~Giddings, Phys. Rev. D {\bf 66}, 044011 (2002);
S.N. Solodukhin, Phys. Lett. {\bf B533}, 153 (2002);
A. Jevicki and J. Thaler, Phys. Rev. D{\bf 66}, 024041 (2002).

\bibitem{ag}  
L. Anchordoqui and H. Goldberg, Phys. Rev. D {\bf 65}, 047502 (2002);
D{\bf 67}, 064010 (2003).

\bibitem{ppch}  
R. Casadio and B. Harms, Int. J. Mod. Phys. A {\bf 17}, 4635 (2002).

\bibitem{ppk}  
K. Cheung, Phys. Rev. D {\bf 66}, 036007 (2002); 
Phys.  Rev. Lett. {\bf 88}, 221602 (2002); 
K. Cheung and Chung-Hsien Chou, Phys. Rev. D {\bf 66}, 036008 (2002).

\bibitem{ppu} Y. Uehara, Mod. Phys. Lett. A {\bf 17}, 1551 (2002).

\bibitem{park} 
Seong Chan Park and H.S. Song, J. Korean Phys. Soc. {\bf 43}, 30 (2003). 

\bibitem{hof} M. Bleicher, S. Hofmann, S. Hossenfelder and H. Stoecker, 
Phys. Lett. B {\bf 548}, 73 (2002); 
S. Hossenfelder, S. Hofmann, M. Bleicher and H. Stoecker, 
Phys. Rev. D {\bf 66}, 101502 (2002);
S. Hossenfelder, B. Koch and M. Bleicher, [arXiv:hep-ph/0507140];
B. Koch, M. Bleicher and S. Hossenfelder, JETP {\bf 0510}, 053 (2005); 
S. Hossenfelder, [arXiv:hep-ph/0510236].

\bibitem{more} 
I. Mocioiu, Y. Nara and I. Sarcevic, Phys. Lett. B{\bf 557}, 87 (2003); 
V. Frolov and D. Stojkovic, Phys. Rev. D {\bf 66}, 084002 (2002); 
{\bf 67}, 084004 (2003); {\bf68}, 064011 (2003); 
D. Ida and S. C. Park, Phys. Rev. D {\bf67}, 064024 (2003); {\bf 69}, 049901 
(E) (2004); 
B. Kol, hep-ph/0207037;
M. Cavagli\`a, S. Das and R. Maartens,
Class. Quant. Grav. {\bf 20}, 1205 (2003);
M. Cavagli\`a and S. Das, Class. Quant. Grav. {\bf 21}, 4511 (2004);
{\bf 20}, L205 (2003);
S. Hosenfelder, Phys. Lett. {\bf B598}, 92 (2004);
A. Ringwald, Fortsch. Phys. {\bf 51}, 830 (2003);
V. Cardoso, E. Berti and M. Cavagli\`a, [arXiv:hep-ph/0505125];
V. Cardoso, M. Cavagli\`a and L. Gualtieri, [arXiv:hep-th/0512002];
M. Cavagli\`a and S. Das, Class. Quant. Grav, {\bf 21}, 4511 (2004);
E. Berti, M. Cavagli\`a and L. Gualtieri, Phys. Rev. D{\bf 69}, 124011 (2004);
R. Godang {\it et al.,} Int.J.Mod.Phys, {\bf A20}, 3409 (2005);
E-J. Ahn and M. Cavagli\`a, Int.Jour.Mod.Phys. {\bf D12}, 1699 (2003).

\bibitem{gram}
T.~Han, G.~D.~Kribs and B.~McElrath, Phys. Rev. Lett. {\bf 90}, 031601 (2003). 

\bibitem{cham} 
A. Chamblin and G. C. Nayak, Phys. Rev. D {\bf 66}, 091901 (2002).

\bibitem{pp4} 
L. A. Anchordoqui, J. L. Feng, H. Goldberg and A. D. Shapere,
Phys. Rev. D {\bf 66}, 103002 (2002); D {\bf 68}, 104025 (2003);
Phys. Lett. {\bf B594}, 363 (2004); 
L. A. Anchordoqui, H. Goldberg and A. D. Shapere,
Phys. Rev. D{\bf 66}, 024033 (2002);
J.~L.~Feng and A.~D.~Shapere, Phys. Rev. Lett. {\bf 88}, 021303 (2002); 
L. A. Anchordoqui, T. Paul, S. Reucroft and J. Swain, Int. J. Mod. Phys.
A {\bf18}, 2229 (2003). 

\bibitem{pp5} 
R.~Emparan, M.~Masip and R.~Rattazzi, Phys. Rev. D {\bf 65}, 064023 (2002).

\bibitem{pp6} 
A.~Ringwald and H.~Tu, Phys.\ Lett.\ B {\bf 525}, 135 (2002);

\bibitem{gravr} 
G. F. Giudice, R. Rattazzi and J. D. Wells, Nucl. Phys. {\bf B544}, 3 (1999); 
L. Vacavant and I. Hinchliffe, J. Phys. G {\bf 27}, 1839 (2001); 
G. C. Nayak, hep-ph/0211395; 
Y. A. Kubyshin, hep-ph/0111027; 
S.~B.~Bae et al., Phys.\ Lett.\ B {\bf 487}, 299 (2000);
S. C. Park, H. S. Song and J. Song, Phys. Rev. D {\bf 63}, 077701 (2001);
K.~Cheung, Phys.\ Rev.\ D {\bf 63}, 056007 (2001).

\bibitem{gravr1} 
W.~D.~Goldberger and M.~B.~Wise,
Phys.\ Rev.\ Lett.\  {\bf 83}, 4922 (1999);
Phys.\ Lett.\ B {\bf 475}, 275 (2000);
Phys.\ Rev.\ D {\bf 60}, 107505 (1999);
C.~Cs\'aki, M.~Graesser, L.~Randall and J.~Terning,
Phys.\ Rev.\ D {\bf 62}, 045015 (2000);
C.~Cs\'aki, M.~L.~Graesser and G.~D.~Kribs,
Phys.\ Rev.\ D {\bf 63}, 065002 (2001);
D {\bf 62}, 067505 (2000);
C.~Cs\'aki, J.~Erlich and J.~Terning,
Phys.\ Rev.\ D {\bf 66}, 064021 (2002);
C.~Cs\'aki, J.~Erlich, T.~J.~Hollowood and Y.~Shirman,
Nucl.\ Phys.\ {\bf B581}, 309 (2000);
E.~A.~Mirabelli, M.~Perelstein and M.~E.~Peskin,
Phys.\ Rev.\ Lett.\  {\bf 82}, 2236 (1999);
D.~Dominici, B.~Grzadkowski, J.~F.~Gunion and M.~Toharia, 
Nucl. Phys. {\bf B671}, 243 (2003);
T.~Han, G.~D.~Kribs and B.~McElrath, Phys.\ Rev.\ D {\bf 64}, 076003 (2001);
M.~Chaichian, A.~Datta, K.~Huitu and Z.~Yu,
Phys.\ Lett.\ {\bf B524}, 161 (2002);
J.~L.~Hewett and T.~G.~Rizzo, J. High Energy Phys. {\bf 08}, 028 (2003); 
A. Chamblin and G.W. Gibbons, Phys. Rev. Lett. {\bf 84}, 1090 (2000);
For a review see G.~D.~Kribs, [arXiv:hep-ph/0110242].

\bibitem{large} 
A. Chamblin, S. W. Hawking and H. S. Reall, 
Phys. Rev. D {\bf 61}, 065007 (2000);
R. Emparan, G. T. Horowitz and R. C. Myers, J. High Energy Phys.
{\bf 01}, 07 (2000); 
N. Dadhich, R. Maartens, P. Papadopoulos and
V. Rezania, Phys. Lett. B {\bf 487}, 1 (2000); 
A. Chamblin, H. Reall, H. Shinkai and
T. Shiromizu, Phys. Rev. D {\bf 63}, 064015 (2001); 
P. Kanti and K. Tamvakis, Phys. Rev. D {\bf 65}, 084010 (2002);
D.J.H. Chung and K. Freese, Phys. Rev. D{\bf 61}, 023511 (2000); 
C. Germani and R. Maartens, Phys. Rev. D {\bf 64}, 124010 (2001); 
I. Giannakis and H. Ren, Phys. Rev. D {\bf 63}, 125017 (2001); 
R. Casadio and L. Mazzacurati, Mod. Phys. Lett. A{\bf 18}, 651 (2003); 
P. Kanti, I. Olasagasti and K. Tamvakis, Phys. Rev. D {\bf 66} (2002) 104026.

\bibitem{susy} 
A. Chamblin, F. Cooper and G. C. Nayak, Phys. Rev. D {\bf 70}, 075018 (2004). 

\bibitem{been} 
W. Beenakker, R. H\"opker, M. Spira and P.M. Zerwas,
Nucl. Phys. {\bf B492}, 51 (1997);
W. Beenakker, H. Kuijf, W.L. van Neerven and J. Smith,
Phys. Rev. D {\bf 40}, 54 (1989).

\bibitem{gray4} 
L. Anchordoqui and H. Goldberg, Phys. Rev. D {\bf 67}, 064010 (2003). 

\bibitem{bv} 
R. Emparan, G. T. Horowitz and R. C. Myers, 
Phys. Rev. Lett. {\bf 85}, 499 (2000).

\bibitem{grayd} P. Kanti and J. March-Russell, 
Phys. Rev. D {\bf 67}, 104109 (2003). 

\bibitem{cooper} 
A. Chamblin, F. Cooper and G. C. Nayak, Phys. Rev. D {\bf 69}, 065010 (2004). 

\bibitem{plasma} 
A. Chamblin, F. Cooper and G.C. Nayak, Phys. Rev. D{\bf 69}, 065010 (2004);
F. Cooper, E. Mottola and G. C. Nayak, 
Phys. Lett. B {\bf 555}, 181 (2003); 
G. C. Nayak {\it et al.}, Nucl. Phys. {\bf A687}, 457 (2001);
R. S. Bhalerao and G. C. Nayak, Phys. Rev. C {\bf 61}, 054907 (2000); 
G. C. Nayak and V. Ravishankar, Phys. Rev. C {\bf 58}, 356 (1998); 
Phys. Rev. D {\bf 55}, 6877 (1997).

\bibitem{YR} 
H. Yoshino and Y. Nambu, Phys. Rev. D{\bf 67}, 024009 (2003);
H. Yoshino and V. S. Rychkov, Phys. Rev. D{\bf 71}, 104028 (2005).

\bibitem{ravindran}
D. Choudhury, S. Majhi and V. Ravindran, JHEP {\bf 0601}, 027 (2006);
P. Mathews, V. Ravindran and K. Sridhar, JHEP {\bf 0408}, 048 (2004);
P. Mathews and V. Ravindran, [arXiv:hep-ph/0507250];
P. Mathews, V. Ravindran, K. Sridhar and W.L. van Neerven,
Nucl. Phys. {\bf B713}, 333 (2005);
S. Lola, P. Mathews, S. Raychaudhuri and K. Sridhar, [arXiv:hep-ph/0010010].

\bibitem{allanach}
B.C. Allanach {\it et al.,} JHEP {\bf 0212}, 039 (2002).

\bibitem{landsberg}
G. Landsberg, Phys. Rev. Lett. {\bf 88}, 181801 (2002);
Eur. Phys. J. {\bf C33}, S927 (2004);
[arXiv:hep-ex/0412028];
N. Akchurin, {\it et al.,} FERMILAB-FN-0752 June 2004.

\bibitem{harris}
C.M. Harris, Ph.D. thesis, [arXiv:hep-ph/0502005];
C.M. Harris {\it et al.,} JHEP {\bf 0505}, 053 (2005);
C.M. Harris, P. Richardson and B.R. Webber, JHEP {\bf 0308}, 033 (2003);
C.M. Harris and P. Kanti, Phys. Lett. {\bf B633}, 106 (2006);
JHEP {\bf 0310}, 014 (2003).

\bibitem{reviews}
P. Kanti, Int. J. Mod. Phys. {\bf A19}, 4899 (2004);
G. Landsberg, [arXiv;hep-ph/0211013];
M. Cavagli\`a, Int. J. Mod. Phys. {\bf A18}, 1843 (2003);
S. Hossenfelder, in {\it Focus on Black Hole Research}, p 155
Nova Science Publishers (2005) [arXiv;hep-ph/0412265];
B. Webber, [arXiv:hep-ph/0511128].

\bibitem{rsvn} V. Ravindran, J. Smith and W. L. van Neerven, 
Nucl. Phys. {\bf B665}, 325 (2003).

\bibitem{others}
R.V. Harlander and W.B. Kilgore, Phys. Rev. Lett.
{\bf 88}, 201801 (2002); 
C. Anastasiou and K. Melnikov, Nucl. Phys. {\bf B646}, 220 (2002).

\bibitem{smith} V. Ravindran, J. Smith and W. L. van Neerven, 
Nucl. Phys. {\bf B634}, 247 (2002).

\bibitem{diff1}
D. de Florian, M. Grazzini and Z. Kunszt, Phys. Rev. Lett. 
{\bf 82}, 5209 (1999);
C.J. Glosser and C.R. Schmidt, JHEP {\bf 0212}, 016 (2002);
J. Smith and W.L van Neerven, Nucl. Phys. {\bf B720}, 182 (2005);
D. de Florian, A. Kulesza and W. Vogelsang, [arXiv:hep-ph/0511205].

\bibitem{lep}
R. Barate {\it et al.,} (ALEPH Collaboration), Phys. Lett. {\bf B495}, 1 (2000);
M. Acciarri {\it et al.,} (L3 Collaboration), Phys. Lett.  {\bf B495}, 18 (2000);
P. Abreu {\it et al.,} (DELPHI Collaboration), Phys. Lett. {\bf B499}, 23 (2001);
G. Abbiendi {\it et al.,} (OPAL Collaboration), Phys. Lett. {\bf B499}, 
38 (2001);
R. Barate {\it et al.,} Phys. Lett. {\bf B565}, 61 (2003).

\bibitem{ghkd}
J.F. Gunion, H.E. Haber, G.L. Kane and S. Dawson, "The Higgs Hunter's Guide",
(Addison-Wesley, Reading, M.A.,1990);
J.F. Gunion, H.E. Haber, G.L. Kane and S. Dawson, 
[arXiv;hep-ph/9302272] Erratum.

\bibitem{wil}
F. Wilczek, Phys. Rev. Lett. {\bf 39}, 1304 (1977);
H. Georgi, S. Glashow, M. Machacek and D. Nanopoulos, Phys. Rev. Lett,
{\bf 40}, 692 (1978);
J. Ellis, M. Gaillard, D. Nanopoulos and C. Sachrajda, Phys. Lett.
{\bf B83}, 339 (1979);
T. Rizzo, Phys. Rev. D{\bf 22}, 178 (1980).

\bibitem{ehsb}
R.K. Ellis, I. Hinchliffe, M. Soldate and J.J. van der Bij, Nucl. Phys.
{\bf B297}, 221 (1988);
U. Baur and E. Glover, Nucl. Phys, {\bf B339}, 38 (1990);
B. Field, S. Dawson and J. Smith, Phys. Rev. D{\bf 69}, 074013 (2004).

\bibitem{gsz}
D. Graudenz, M. Spira and P. Zerwas, Phys. Rev. Lett. {\bf 70}, 1372 (1993);
M. Spira, A. Djouadi, D. Graudenz and P. Zerwas, Phys. Lett. {\bf B318}, 347
(1993); Nucl. Phys. {\bf B453}, 17 (1995).

\bibitem{dkosz}
V. Del Duca, W. Kilgore, C. Oleari, C. Schmidt and D. Zeppenfeld,
Phys. Rev. Lett. {\bf 87}, 122001 (2001); Nucl. Phys. {\bf B616}, 367 (2001).

\bibitem{dawson}
S. Dawson, Nucl. Phys. {\bf B359}, 283 (1991);
A. Djouadi, M. Spira and P. Zerwas, Phys. Lett. {\bf B264}, 440 (1991).

\bibitem{kls}
M. Kr\"amer, E. Laenen and M. Spira, Nucl. Phys. {\bf B511}, 523 (1998).

\bibitem{kdr}
R.P. Kauffman, S.V. Desai and D. Risal, Phys. Rev. D{\bf 55}, 4005 (1997);
Phys. Rev. D{\bf 58}, 11901 (1998), 
Erratum.

\bibitem{schmidt}
C.R. Schmidt, Phys. Lett. {\bf B413}, 391 (1997).

\bibitem{fgk}
D. de Florian, M. Grazzini and Z. Kunszt, Phys. Rev. Lett. {\bf 82}, 
5209 (1999);
S. Catani, D. de Florian and M. Grazzini, JHEP {\bf 0201}, 015 (2002).

\bibitem{harland}
R.V. Harlander, Phys. Lett. {\bf B492}, 74 (2000).

\bibitem{rsvn2} V. Ravindran, J. Smith and W.L. van Neerven,
Nucl. Phys. {\bf B704}, 332 (2005). 

\bibitem{haki1}
S. Catani, D. de Florian and M. Grazzini, JHEP {\bf 0105}, 025 (2001); 
R.V. Harlander and W.B. Kilgore, Phys. Rev. {\bf D64}, 013015 (2001); 
S. Catani, D. de Florian, M. Grazzini and P. Nason, JHEP {\bf 07}, 028 (2003);
S. Moch and A. Vogt, Phys. Lett. {\bf B631}, 48 (2005);
V. Ravindran, [arXiv:hep-ph/0512249].

\bibitem{cks}
K.G. Chetyrkin, B.A. Kniehl and M. Steinhauser, Phys. Rev. Lett. {\bf 79},
2184 (1997).

\bibitem{klzu}
H. Kluberg-Stern and J.B. Zuber, Phys. Rev. {\bf D12}, 467 (1975);
V.P. Spiridonov and K.G. Chetyrkin, Yad. Fiz. {\bf 47}, 818 (1988).

\bibitem{mvv}
S. Moch, J.A.M. Vermaseren and A. Vogt, Nucl. Phys. {\bf B688}, 101 (2004);
A. Vogt, S. Moch and J.A.M. Vermaseren, Nucl. Phys. {\bf B691}, 129 (2004).

\bibitem{larin2}
S. A. Larin, T. van Ritbergen and J.A.M. Vermaseren, Nucl. Phys. {\bf B427},
41 (1994);
S. A. Larin, P. Nogueira, T. van Ritbergen and J.A.M. Vermaseren, 
Nucl. Phys. {\bf B492}, 338 (1997);
A. Ret\'ey and J.A.M. Vermaseren, Nucl. Phys. {\bf B604}, 281 (2001).

\bibitem{nevo}
W.L. van Neerven and A. Vogt, Nucl. Phys. {\bf B568}, 263 (2000);
Nucl. Phys. {\bf B588}, 345 (2000); Phys. Lett. {\bf B490}, 111 (2000).

\bibitem{mrst01}
A.D. Martin, R.G. Roberts, W.J. Stirling and R.S. Thorne, Eur. Phys. J.
{\bf C23}, 73 (2002).

\bibitem{mrst02}
A.D. Martin, R.G. Roberts, W.J. Stirling and R.S. Thorne, Phys. Lett.
{\bf B531}, 216 (2002).

\bibitem{chkn}
K.G. Chetyrkin, B.A. Kniehl and M. Steinhauser, Phys. Rev. Lett. {\bf 79}, 
2184 (1997).

\bibitem{lai}
J. Pumplin et al., JHEP {\bf 0207}, 102 (2002).

\bibitem{bc}
E.L. Berger and J. Campbell, Phys Rev. {\bf D70}, 073011 (2004).
\end{thebibliography}
\end{document}